\documentclass[12pt,preprint]{aastex}

\begin{document}

\title{Late Light Curves of Type Ia Supernovae}

\author{P.A. Milne\footnote{NAS/NRC Resident Research Associate.}}
\affil{Naval Research Laboratory, Code 7650, Washington, DC 20375\\}
\author{L.-S. The, M.D. Leising}
\affil{Department of Physics and Astronomy\\
    Clemson,S.C., 29634-0978}


\begin{abstract}
We extend earlier efforts to determine whether the late 
(t$\geq$60d) light-curves
of type Ia SNe are better explained by 
the escape of positrons from the ejecta 
or by the complete deposition of positron kinetic
energy in a trapping 
magnetic field.
We refine our selection of
Ia SNe, using those that have 
extensive BVRI photometry 35 days or more after maximum light.
Assuming all SNe within a given $\Delta$m$_{15}$(B) range form a 
distinct sub-class, we fit a combined light-curve for 
all class members with a variety of models. 
We improve our previous calculations of energy deposition rates by including
the transport of the comptonized electrons. Their non-local and
time-dependent energy deposition produces a correction 
of as much as $\leq$0.10$^m$ for Chandrasekhar models 
and $\leq$0.18$^m$ for 
sub-Chandrasekhar models. 

We find that applying a filter efficiency correction, derived
from measured spectra, to B, V, R,
and I light-curves after day 50 can produce a consistent bolometric
light-curve.
The V band is an accurate indicator of total emission
in the 
3500$\AA$ - 9700$\AA$ range, with a constant fraction ($\sim$25\%)
appearing in the V band after day 50.
This suggests that the V band 
scales with the bolometric luminosity, and that the deposited energy
is instantaneously recycled into optical emission 
during this epoch. Varying bolometric corrections
for the other bands are derived. 
We see significant evolution of the colors of SNe Ia  
between day 50 and day 170.  We suggest that 
this may be due to the transition from spectra dominated by 
emission lines from 
the radioactive nucleus, $^{56}$Co, to those 
from the stable daughter nucleus, $^{56}$Fe. 
We show that the B, V, R, and I band light-curves 
of SNe Ia after t$\geq$60d can be completely explained
with energy deposition from $^{56}$Co decay 
photons and positrons if substantial positron escape occurs.

\end{abstract}

\section{Introduction}

Type Ia supernovae (SNe) are an integral part of many current 
astrophysical investigations. Their luminosities briefly 
rival entire galaxies, making them useful as distance 
indicators to large $z$ (Riess et al. 1998;
Perlmutter et al. 1999). Understanding of their 
contributions to
galactic chemical evolution is necessary to sort out the
star formation history of various other 
stellar populations 
in the Galaxy (Timmes et al. 1996). The escape of gamma 
rays from their ejecta may produce a significant 
portion of the diffuse gamma ray background in the 
400 - 2000 keV energy range (Watanabe et al. 1999). The 
escape of positrons from their ejecta may be large 
enough to explain a majority of the 511 keV annihilation 
radiation measured by the CGRO/OSSE instrument 
(Chan \& Lingenfelter 1993; Milne, The \& Leising 1999, hereafter 
MTL). 
All of these investigations rely upon a basic understanding
of the SN Ia event. 

Type Ia SNe are widely-believed to be the thermonuclear explosion of a 
mass-accreting carbon-oxygen white dwarf (WD). 
The conversion of carbon and oxygen to 
iron-peak and intermediate-mass elements releases a large amount 
of energy initially, and subsequently the decay $^{56}$Ni $\rightarrow$ 
$^{56}$Co $\rightarrow$ $^{56}$Fe powers the supernova
through gamma- and X-ray photons and positrons.\footnote{Although 
neutrinos are also created 
in these decays, the lower neutrino opacity makes neutrinos an insignificant
contributor to the energy deposition rate during the epoch of 
interest in this work. See Nadyozhin (1994) for major features 
of this decay chain.}
The deposition and diffusion of this energy produces the 
optical emission that defines the Type Ia SN. The observational support 
for this paradigm is typified by the ability of a standard carbon-oxygen WD 
model, W7, to reproduce the light curves and spectra of the SN Ia 1981B 
(Branch et al. 1985). 

During the earliest epoch after the SN explosion, 
the energy released in the $^{56}$Ni $\rightarrow$
$^{56}$Co $\rightarrow$ $^{56}$Fe decay is 
efficiently trapped by the SN ejecta. The deposition of this energy creates 
energetic electrons, which thermalize (primarily via 
ionization and excitation of atoms)
and recombine. The optical photons created in 
the recombinations diffuse to the surface, and escape the SN ejecta. The 
emerging optical spectrum is dictated by the nature of the diffusion of the
optical 
photons. The expansion of the SN ejecta lowers the column density to 
the surface, decreasing both the time for optical photons to
diffuse outward and the efficiency of trapping of gamma-rays 
and energetic positrons and electrons.
The SN thus transitions from an 
epoch during which the energy deposition is essentially 
complete and instantaneous and 
the emission depends upon the diffusion of optical photons (a 
``diffusion-dominated" epoch) 
to an epoch during which the diffusion time-scale is 
negligibly short and the emerging emission depends upon the transport of 
the decay products (a ``deposition-dominated" epoch). 

Observations of SNe during these two epochs probe different characteristics
of the 
SN explosion. 
As will be described in the next section, observations of SNe Ia during 
both epochs have led to considerable advances in the understanding of the 
SN event. This work concentrates upon late emission. It follows a previous 
work (Milne, The \& Leising 1999) which also investigated late emission,   
fitting model-generated energy deposition rates to 
photometry of 10 SNe Ia. That work treated the 50$^{d}$ -200$^{d}$ 
time period in an approximate fashion, concentrating upon the 200$^{d}$ -
1000$^{d}$ emission and will be referenced for discussions relating to 
gamma-ray, X-ray and 
positron transport and the yield of escaping positrons. This 
work performs similar comparisons, but 
concentrates more on the 50$^{d}$ -200$^{d}$ period, 
which spans the
transition from diffusion-dominated emission to deposition-dominated
emission.\footnote{In this work, all SNe Ia will be assumed to rise to
peak luminosity in 18 days.}
In addition, this paper treats the information that is 
available from the R and I band observations of SNe Ia, as well as from late
spectra.  We compare model-generated energy deposition 
rates with B,V,R, and I band photometry for a large collection of 
well-observed SNe Ia, before and after creating bolometric corrections 
from a collection of late SN Ia spectra. 
Through these comparisons, we address four specific 
questions: 1) Is there any order to the late light curves of SNe Ia? 
 2) Is there observational support of the suggestion 
that positrons escape the 
SN ejecta? 3) Do the late light curves afford any insight into the 
correct explosion scenario(s)? 4) Are the late light curves of SNe Ia
suitably 
explained by the decay products of the $^{56}$Ni $\rightarrow$
$^{56}$Co $\rightarrow$ $^{56}$Fe without the inclusion of additional 
sources of energy deposition.

\section{Physics of SN Ia Emission}

\subsection{Emission during the first 50 days}

SN light curves peak in brightness and fade on time-scales related to
the lifetimes of $^{56}$Ni ($\tau$=8.8$^{d}$) and 
$^{56}$Co ($\tau$=111$^{d}$).
By 100 days after the explosion, SN Ia emission has faded 
to less than 4\% of its maximum luminosity. Largely because 
of the dimming of the emission, SNe Ia are better studied 
during the first 50 days than during later epochs. As a result, 
most investigations address issues relating to the diffusion-dominated 
epoch.  
Once thought to be a homogeneous class, more and better 
observations have revealed that intrinsic differences 
exist among SNe Ia (Phillips et al. 1987, 
Filippenko et al. 1992, 1992b, Suntzeff 1996). 
These differences are evident in the time evolution of the 
spectra, the shape of the light curves, and in the 
absolute magnitude of the emission at peak luminosity. 
Inhomogeneity at the level observed has necessitated the 
enlargement of the paradigm from a single, standard model 
to families of models. 

The level of understanding of this inhomogeneity has progressed 
on many fronts over the last decade. Observationally, ``twins" 
have been found for both anomalously bright, and anomalously 
faint SNe Ia, solidifying the existence of a large range
(and perhaps a continuum) of phenomena.\footnote{Li et al. 2000
concluded that only 64\% of all SNe Ia are normally-luminous, 
with $\sim$16\% sub-luminous and 
$\sim$20\% super-luminous.} The very existence of 
inhomogeneity has altered the investigations of explosion scenarios.
Four explosion scenarios dominate the present paradigm. The first 
consists of a carbon-oxygen white dwarf near the Chandrasekhar
mass accreting hydrogen or helium from a binary companion until it reaches 
a mass at which the core carbon ignites. If the resulting burning front
accelerates to become a detonation in the outer layers of the 
WD, a ``delayed detonation" results. If the burning front remains 
sub-sonic, the result is a ``deflagration". These Chandrasekhar mass 
(CM) scenarios account for 
inhomogeneity by variations in the propagation of the burning
front due to density and/or 
compositional differences in the progenitor carbon-oxygen  WD. 
The second scenario consists of a lower mass carbon-oxygen WD accreting a 
helium shell, which becomes thick enough to produce a helium shell 
detonation. This, in turn, triggers central carbon ignition. In this 
sub-Chandrasekhar (SC) mass scenario, inhomogeneity is due to the 
different nucleosynthesis that results when the progenitor mass varies 
from 0.65 - 1.1 M$_{\odot}$. The third scenario merges two 
carbon-oxygen white dwarfs, with the more massive white dwarf 
accreting the companion. As with the Chandrasekhar
mass scenario, central carbon ignition results. In this case, the 
accreted envelope is carbon. The range of masses of these 
explosions have been suggested to vary from 1.2 - 1.8 M$_{\odot}$, 
with a roughly constant nickel yield. 
This scenario is also referred to as a ``double degenerate" 
explosion. 
The fourth scenario is not a thermonuclear explosion at all, but an 
accretion-induced collapse (AIC) of a white dwarf (either CO or ONeMg). 
It has been argued that in some cases, electron capture within an 
accreting white dwarf may lead to a collapse 
 rather than central carbon ignition (as with Types II/Ib/Ic
SNe). These events eject less total mass and less nickel 
than the other scenarios, 
and have been suggested to explain sub-luminous SNe Ia. 
Families of SN models have been 
developed within these scenarios (including variations of
these scenarios such as pulsed delayed detonations). Using these 
models, many authors have demonstrated that compositional and 
kinematic differences within families of these SN models 
can roughly simulate spectral variations in observed SNe.  
These differences are seen and simulated near peak luminosity 
(Mazzali et al. 1992, 1995, 1997, H\H{o}flich et al. 1996, 
Jeffery et al. 1992, Baron et al. 1996), and during the 
later, nebular epoch (Ruiz- Lapuente et al. 1992, 1996, 
Liu et al. 1997a, 1997b, 1998, Bowers et al. 1997). 

The existence of inhomogeneity affects the use of 
SNe Ia as distance indicators at the level
currently employed.
There is an apparent absolute luminosity/peak-width relationship 
inferred from distance estimates to the host galaxies of nearby 
SNe Ia, quantified variously as  
$\Delta$m$_{15}$(B) (Phillips 1993), MLCS (Riess et al. 1996), 
and stretch (s) (Perlmutter et al. 1997). The ability to 
theoretically explain the luminosity/peak-width relationship 
is critical for relating well-observed, nearby SNe to SNe 
at the cosmological distances at which SNe Ia are used as 
distance indicators. This task  
is difficult because calculating the transport of optical photons through SN
ejecta is a very complicated procedure. The ejecta is 
constantly evolving in both density and temperature. A  
photon crossing this non-equilibrated ejecta is 
red- (or blue-) shifted relative to local matter, 
wreaking havoc with the line-dominated 
opacity. Further complicating matters are the uncertainties of the 
cross sections of many relevant interactions. Despite these difficulties, 
detailed studies have been performed. H\H{o}flich (1995) 
fit the B,V,R,I band observations of the well-observed SN 1994D with a 
variety of Chandrasekhar mass models, demonstrating 
a high level of discrimination between 
models. H\H{o}flich \& Khokhlov (1996) then generated B-M band 
light curves for Chandrasekhar mass and sub-Chandrasekhar models, 
and fit B,V,R,I band observations of 26 SNe Ia. One conclusion from that 
study is that sub-Chandrasekhar models appear too blue at peak to explain 
observations of sub-luminous SNe Ia. The same conclusion was reached 
by Nugent et al. (1997). Pinto \& Eastman (1996, 2000a,b,c) 
investigated the influence of progenitor mass, nickel mass, nickel
distribution, explosion energy and opacity upon the bolometric light
curves. They then concentrated upon the ability of
Chandrasekhar mass models to reproduce the B,V,R luminosity/peak-width
relation 
in the range 
0.85 $\leq$ $\Delta$m$_{15}$(B) $\leq$ 1.75. A significant 
achievement of those efforts is the agreement between their 
calculated 18$^{d}$ -20$^{d}$ rise-times of the B and V peak luminosity, 
with the observed rise-times of 
SNe Ia (Riess et al. 1999a,b; Aldering, Knop \& Nugent 2000).\footnote{The 
models shown in H\H{o}flich (1995) and H\H{o}flich \& Khokhlov 
(1996) rise to V peak in 13$^{d}$ -15$^{d}$. Recent simulations  
by H\H{o}flich et al. (1998) show that the 
V peak can be delayed by $\sim$3 days depending upon the C/O ratio, but 
fits of these models to observed SNe (equivalent to the previous works) 
have not been published.}  

Collectively, these investigations have reached the loose consensus that 
the Chandrasekhar mass 
explosion scenario is the favored scenario to account for the range of
optical 
observations. Whether any subset of SN Ia events occur as sub-Chandrasekhar,
merger or AIC explosions remains 
unclear. In this work, we simulate light curves for 
models representative of all four scenarios, 
and demonstrate general results of late light curve studies.

\subsection{Emission after 50 days}

Neither H\H{o}flich \& Khokhlov (1996) nor Pinto \& Eastman carried their 
photometric simulations 
beyond 120$^{d}$, both groups suggesting that their simulations become 
inadequate at late epochs. As described in MTL, 
during the time interval of interest in this work, the energy
deposition is dominated by interactions involving the decay products of the
$^{56}$Co $\rightarrow$ $^{56}$Fe decay; gamma-ray photons and
positrons. The photons possess $\sim$30 times more energy per
decay than the positrons, but are more penetrating. This leads to
the photons initially dominating the energy deposition but 
transitioning at later times to positron dominance. The photon
transport was performed with a Monte Carlo algorithm adopting the
prescription of Podznyakov, Sobol, \& Sunyaev (1983). A detailed
description of the Monte Carlo algorithm is given in The et al. (1990), 
with its application to SNe Ia light curves in Burrows \& The
(1990).\footnote{The resulting gamma-ray light curves agree with
H\H{o}flich (1995) and H\H{o}flich et al. (1996) to within 5\%.} The
dominant interaction is Compton scattering,
which produces energetic electrons. These secondary electrons have a
mean energy typically 300 keV at 100$^{d}$. MTL assigned these
electrons zero lifetimes, depositing the energy {\it in-situ}.
This work improves on that by including the transport of the
secondary electrons with the same algorithm employed for the positron
transport, but using M\H{o}ller scattering rather than Bhabha scattering.
The effect of the secondary-electron transport is small but non-negligible;
in the Chandrasekhar mass models with radial positron escape and low
ionization, the
light curves are fainter by $\leq$ 0.10$^{m}$; for low-mass
sub-Chandrasekhar models, the
effect reaches 0.18$^{m}$.

Positron transport depends upon the nature of the 
magnetic field. Three scenarios have been suggested to model the 
magnetic field (Ruiz-Lapuente \& Spruit 1998, hereafter RLS). 
The first suggests that the field is too weak to 
confine positrons, and positrons follow straight line 
trajectories, with a fraction escaping the ejecta (Colgate, Petschek \& 
Kriese 1980). 
The second suggests a stronger field that 
confines positrons, but with the field lines radially-combed by the 
homologous expansion. The positrons 
spiral along these radial field lines with their pitch-angles
decreasing due to the field gradient (beaming), with a fraction 
escaping the ejecta (Chan \& Lingenfelter 1993). The third scenario 
suggests a strong field that is turbulently disordered such that positrons 
mirror frequently with no net transport (Axelrod 1980). 
Positrons may survive non-thermally at late times, but none escape. 
Colgate, et al. (1980) argued that the 
first two situations are equivalent.
Simulations by MTL found this to be approximately true for most 
SN models. The first two scenarios will herein be 
referred to as the ``radial" 
scenario, the third will be referred to as the ``trapping" scenario. 
The differing positron escape leads to 
differences in model energy deposition rates, which makes late 
observations of SNe Ia a probe of the photon and 
positron transport. 

The positron and secondary electron transport was performed with a Monte
Carlo algorithm, as explained in MTL. The dominant energy loss mechanisms
are ionization and excitation of bound electrons for low levels of
ionization, and plasma excitation for higher ionization. The model-generated
energy deposition rate for the sub-Chandrasekhar model, HED8, is shown in
Figure 1. The
dashed line (D) assumes instantaneous deposition of all decay energy, an
assumption that is invalid due to gamma-ray photon escape by 30$^{d}$. The
dotted line (G) uses the results of the gamma-ray/secondary electron
deposition only and assumes no deposition of positron kinetic energy.
By 100$^{d}$, the deposition of positron kinetic energy is an important
contributor to the total energy deposition rate, but the positron
lifetimes are short enough that all positron curves approximate
instantaneous, {\it in-situ} deposition of positron energy (In). At
later times, densities in the ejecta are low enough to allow escape in
the radial scenario (dark shading, R) and non-thermal survival in the
trapping scenario (light shading, T). Both scenarios were calculated for
a range of ionizations, from 1\% of all nuclei being singly ionized, to
all nuclei being triply ionized.
Positron escape removes energy from the ejecta, leading to
fainter light curves than {\it in-situ} deposition. Lower ionization
permits more positron escape and thus is the fainter edge of the radial
curve. A trapping field permits large non-thermal lifetimes, but no escape.
This leads to storage of energy and a late light curve that is brighter than
{\it in-situ} deposition. For this scenario, low ionization leads to a
larger energy storage and the brighter edge of the trapping 
curve.\footnote{The normally-luminous sub-Chandrasekhar model shown allows a
relatively large escape
fraction for the radial scenario compared to normally-luminous Chandrasekhar
mass models.
Nonetheless, the basic features shown in Figure 1 are characteristic of
all the SN Ia models simulated.}

\subsection{Model/Observation Comparisons}

The ideal model/observation comparison would entail an
explicit NLTE determination of photon and positron energy deposition
rates and from these rates the generation of optical/IR spectra. These
spectra would then be compared with a sequence of observed SN Ia
spectra, comparing the evolution of the total flux from early to very
late epochs. This ideal
situation is far from being realized, both computationally and
observationally. Falling short of that solution, an alternative
approach has been employed in numerous works;
fitting energy deposition rates to multi-band photometry or
estimates of the uvoir bolometric luminosity. 
Two issues must be addressed for these 
comparisons to be meaningful, photon diffusion delay and color evolution.
Photon diffusion 
delay refers to the portion of the 
time delay between the energy deposition (from the scattering of 
gamma- and X-ray photons and slowing of positrons) and the production of the
optical 
light that can be seen by an external observer that is due to the diffusion
of 
optical photons out of the ejecta. This diffusion delay  determines the
shape of 
the SN light curve peak at early times, but decreases to negligible values
at late times. 
Thus, at late epochs, energy deposition is instantaneously recycled 
into optical emission. 
In previous studies of late light curves of SNe Ia, two approaches have 
been used to account for photon diffusion delay. 
Studies that attempt to fit observations 
continuously from early to late epochs derive this time delay as a function
of explosion 
epoch. The alternative approach is to study only the late epochs and assume
instantaneous recycling. 
Observationally, the late epoch begins when the emitted spectrum transitions
from continuum emission to
(forbidden) nebular lines. Color evolution refers to the changing fraction
of the total optical 
flux that a given photometric band samples due to the evolution of the
spectrum. Again, two 
different approaches have been employed. The first approach creates
bolometric light curves 
from the photometry (applying weighting factors derived from spectra) and
compares these light 
curves to model-generated energy deposition rates.\footnote{Contardo et al.
(2000) 
discuss the 
issues related to generating bolometric light curves from band photometry.} 
The second approach determines the epoch after 
which a given photometric band scales with the total optical flux (the epoch
is determined from 
spectra), and compares this band photometry to model-generated energy
deposition rates.

Two groups that carried their investigations to late epochs 
have treated photon diffusion in their comparisons. 
Colgate, Petschek and Kriese (1980) fit Monte Carlo model simulations of the
energy 
deposition rate to B band data from SNe 1937C and 1972E. 
That work assumed a grey opacity for both photon and positron transport  
and a single-zone model. They concluded that positron escape was required 
to explain the late light curves. Cappellaro et al. (1998) (hereafter CAPP) 
and Salvo et al. (2001) transported gamma-ray photons, positrons and 
optical photons through W7-like models (scaled to various masses) 
with a Monte Carlo code and fit the energy deposition rates to V band 
data from SNe 1991T, 1994D, 1992A, 1993L, 1996X and 1991bg.\footnote{Both 
Colgate et al. (1980) and CAPP 
transported the positrons with routines that were produced for 
photon transport. The positrons were given a speed $c$.}
That work assumed a grey opacity for photons and 
varied the positron opacity to fit the late light curves. They  
concluded that positron escape was suggested for some, but not all SNe. 

RLS transported gamma-ray photons and positrons 
through Chandrasekhar mass and sub-Chandrasekhar models without 
treating photon diffusion delays, and fit the energy deposition rates 
to bolometric light curves for the SNe 1972E, 1992A and 1991bg.\footnote{The
bolometric light curves were generated from photometry for SNe
1992A and
1991bg. For SN 1972E, the bolometric light curve was generated from a 
series of optical spectra.} 
That work concentrated upon very late times ($\geq$ 100$^{d}$)
at which time the photon diffusion delay time-scale is assumed to be short.
They concluded that normally-luminous SNe Ia transition from a 
positron-trapping phase to a positron-escape phase after 
450$^{d}$, and can be fit 
with Chandrasekhar mass models. They further concluded that
sub-Chandrasekhar models best 
describe the sub-luminous SNe Ia and feature positron escape as early as 
150$^{d}$. MTL transported gamma-ray photons and positrons
through Chandrasekhar mass and sub-Chandrasekhar models, and fit the energy
deposition rates
to bolometric, V band and B band light curves for 10 SNe, including the 
SNe studied by the previously mentioned groups. That work also concentrated 
on very late times and did not treat photon diffusion delays. 
They concluded that 
both Chandrasekhar mass and sub-Chandrasekhar models can explain normally-
and super- luminous 
SNe Ia, with positron escape suggested to be consistent with seven of the
eight 
SNe. They further concluded that none of the models tested could 
explain the sub-luminous SN 1991bg, a result in agreement with CAPP, but in 
disagreement with RLS.

This work does not simulate photon diffusion delays, comparing
model-generated 
energy deposition rates to multi-band photometry as though the delays 
were negligible. This approach uses the comparisons to determine the onset
of 
instantaneous recycling. Similarly, the initial comparisons of the 
model-generated energy deposition rates to each of the B,V,R and I band 
data-sets assume no color evolution. The differences between these 
comparisons and the later comparisons that incorporate spectrally-derived 
bolometric corrections afford a measure of the importance of color 
evolution at late epochs.

\section{SN Ia Observations}

To perform model-generated energy deposition rate/SN light curve 
comparisons, a large collection of SNe Ia observations 
have been compiled. 
The principle sources of B,V,R,I band photometry are the 29 
SNe Ia observed by Hamuy et al. (1996), and the 22 SNe Ia observed by 
Riess et al. (1999). MTL fit the 10 SNe Ia best observed to very late times.
Of those ten SNe, only six are included in this study 
(as
explained below). These SNe have been  
supplemented with observations of SN 1995D by Sadakane et al. (1996), 
SN 1996X by Salvo et al. (2001), SN 1997cn by Turatto et al. (1998),
 SN 1998bu by Jha et al. (1999), Suntzeff et al. (1999), 
Garnavich et al. (2000), Cappellaro et al. (2001),
 and SN 1998de by Modjaz et al. (2000). The SNe Ia have been divided 
into luminosity sub-classes (normally-, sub- and super-luminous). 
Hereafter these categories will be denoted in figures by {\bf N, sb \& SP}.
The categorization is based upon the $\Delta$m$_{15}$(B) (Phillips 1993) 
values of the 
SNe. Numerous empirical relations have been derived for SNe Ia, 
plotting a given variable versus the $\Delta$m$_{15}$(B) parameter. 
Shown in Figure 2 are the maximum B,V and I band absolute magnitudes 
for 21 of the 22 SNe Ia analyzed in this study combined with 18 others from 
Hamuy et al. 1996.\footnote{This figure only included the SNe with 
well-defined peak shapes.} Early studies assumed a linear 
relationship between these variables for all SNe Ia. 
A larger sampling has suggested 
that all these relations deviate from linearity at 
$\Delta$m$_{15}$(B) $\sim$ 1.6. It is unclear whether 
this deviation is best explained 
by a continuous quadratic function applicable to all SNe Ia (Phillips 1999),
or by two different linear relations, one for each sub-class of SNe Ia.
Regardless of the interpretation of this break, the normally-/sub-luminous 
cut has been 
set at 1.60. By that definition, SN 1993H is considered sub-luminous and 
SN 1992A is considered normally-luminous. 
The normally-/super-luminous  
cut ($\Delta$m$_{15}$(B)=0.95) was chosen such that the 
well-observed SN 1991T (which showed spectral deviations from normal SNe Ia)
was considered super-luminous. It is important to note that the 
absence of a template SN with a wider peak than SN 1991T hinders the 
accurate estimation of $\Delta$m$_{15}$(B) for super-luminous SNe. 
As will be shown, the exact location of the normally-/super-luminous
cut is unimportant for this work, but the normally-/sub-luminous 
cut does differentiate two distinct late light curve sub-classes.

From 64 observed SNe Ia, 22 were included in this study. To be included, a 
SN had to be observed at least once after 85$^{d}$ with at least one 
observation within $\pm$20 of the normalization epoch 
(explained below), chosen to be 65$^{d}$ 
post-explosion. Additionally, the SN had to be first observed no later than 
one week post-B maximum (a requirement relaxed for the SN 1992K due to 
the undersampling of the {\bf sb} sub-class). The SNe Ia which meet this 
criteria are; {\bf N}:90N,90O,92A,92al,93ag,94D,95D,95E,95al,96X,98bu, 
{\bf SP}:91T,91ag,92bc,94ae,95bd, {\bf sb}:86G,91bg,92K,93H,97cn,98de. 
Four SNe Ia from MTL were not included in this study. SNe 1937C and 1972E 
were not well-observed with multi-band photometry and were excluded. SN 
1993L was discovered more than a week post-peak and was 
excluded. The late photometry of SN 1989B is being re-analyzed and thus 
that SN was excluded.\footnote{The SN has been identified as a light echo 
candidate (Boffi et al. 1999). As published, the late light curve of 
SN 1989B is more than 1$^{m}$ brighter than the suggested models and is 
the singular exception to the trends seen in the other SNe.} All three 
sub-classes are better sampled from 50$^{d}$ -150$^{d}$ than after
150$^{d}$.

The phenomenon studied in this work is the {\it evolution} 
of the light curves at 
late epochs. For this reason, a relative magnitude scale has been 
employed rather than an absolute or apparent magnitude scale. All SNe have 
been normalized to have zero magnitude at 65$^{d}$ post-explosion 
(assuming an 18$^{d}$ rise-time for all sub-classes). The initial 
motivation for the 65$^{d}$ normalization date was based upon the 
suggestion in Pinto \& Eastman (2000) that the photon diffusion 
time-scale becomes negligibly short at roughly that time. As the next 
section shows, a high level of homogeneity exists within the multi-band 
photometry of each sub-class after 65$^{d}$. Additionally, the greater 
number of 45$^{d}$ -85$^{d}$ observations relative to later epochs makes 
the normalization algorithm 
perform better at early times. These two factors led us to 
retain the 65$^{d}$ normalization date for all data-sets, although the 
model-generated energy deposition rates do not fit the band photometry 
at 65$^{d}$ for any 
band/sub-class combination.  The normalization has been performed by 
linear interpolation of all data between 45$^{d}$ and 85$^{d}$. For two 
SNe, only a single observation existed in that interval. For those SNe, 
the latest pre-45$^{d}$ observation was used to perform the interpolation.  

By treating all SNe within a given $\Delta$m$_{15}$(B) range as a single
object, 
we are assuming homogeneity within each luminosity sub-class. The amount of
scatter about a 
template gives some measure of the quality of that assumption, but numerous 
sources of systematic errors exist to cloud that interpretation. Each SN 
light curve is measured relative to the host galaxy's background light,
which
is unique to that SN. If the SN is near the detectability limit for that 
observation, the influence of the background subtraction increases in 
importance. The irregularity of the sampling means that highly uncertain 
observations are mixed throughout every epoch of the late light curves. 
Similarly, different telescopes, filters 
 and detectors were used to produce this 
data-set, introducing additional systematic differences between individual 
data-points. As was shown by Suntzeff (1999), observations of the same SN by
the same observer at the same site with the same class telescope can 
yield different photometry. The differences were found to be in excess of
the 
quoted uncertainties. Suntzeff showed that in this optimal case, the 
``instrumental uncertainty" was on the order of 0.06$^{m}$ for the V band 
during the 60$^{d}$ -75$^{d}$ epoch. For the case of this mixed 
data-set, the uncertainties are likely to be larger. 
Errors in this work 
were also introduced by the fitting algorithm, which fit the 
individual light curves to the estimated 65$^{d}$ magnitude. 

All errors 
discussed above have been related to estimates of the relative magnitude. As
SNe are always detected at some time after the explosion, the explosion 
date is also somewhat uncertain. Two factors may influence uncertainties 
in the explosion date; 1) uncertainty of a sub-class rise-time to peak, 
and 2) uncertainty of the peak date. For this work, all SNe were assumed 
to rise to peak B magnitude in 18$^{d}$ regardless of the luminosity 
sub-class. To minimize the effect of peak date uncertainty, only SNe first 
observed no later than a week after maximum light were used 
(with the exception of SN 1992K). The 
determination of the peak date depends upon template fitting, so peak date 
uncertainties are only relative to the template used. SNe discovered 
before t$_{B}$(max) have peak date uncertainties typically of $\pm$1$^{d}$. 
SNe discovered post-peak have larger uncertainties. Nonetheless, as the 
time-scales involved in this work are on the order of tens of days, 
uncertainties on the order of a few days will not significantly alter these 
results.

\section{Results}

\subsection{V Band}

The 22 SNe included in this study are shown fit (by eye) with the 
delayed-detonation, Chandrasekhar mass model DD23C in Figure 3.
All 22 SNe have  been normalized at 65$^{d}$. Three tendencies
are apparent in the data. First, within  each sub-class the
data show remarkable homogeneity. The sub-classes are  DEFINED
by differences in the early light curve shape, but after
65$^{d}$,  there is no evidence of variations. Second, the
normally-luminous and super-luminous data appear to
evolve similarly, while the sub-luminous sub-class  continues
to fall more steeply (recall that a steep decline from peak 
DEFINES this sub-class). Third, the separation between the
normally-/super-luminous  and the sub-luminous sub-classes far
exceeds the scatter within the  sub-classes. This argues
against a continuous transition between the  sub-luminous and
normally-luminous sub-classes. Unfortunately, the sub-luminous
sub-class is   under-sampled, so many more sub-luminous SNe
need to be observed before  addressing whether there exists an
absolute separation of sub-classes. 

After $\sim$80$^{d}$ the 
normally-luminous and 
super-luminous sub-classes appear to be explainable by this model, assuming 
radial escape. By contrast, the
sub-luminous sub-class cannot be explained by either magnetic field
scenario. 
To improve the visualization of the 40$^{d}$ -120$^{d}$ 
normally-luminous and super-luminous data, in the inset of Figure 3 
we show the data as residuals relative to the model curve. The 
similarities between 
the normally-luminous and super-luminous data-sets is apparent. 
Also apparent in the inset are the failure of the 
model light curves to fit the V band data in detail until 
after $\sim$120$^{d}$. These results are not specific to 
the delayed-detonation, Chandrasekhar mass model, DD23C. 
Shown in Figure 4 are six 
models suggested to explain normally-luminous and super-luminous SNe, 
shown in the ``delta magnitude" format.\footnote{The delta magnitude format 
shows the residuals of
the data and model-generated light curves to the instantaneous
deposition
approximation, in units of magnitude.} The models are 
defined in Table 1. All model light curves show the same fundamental 
structure, with the models fitting the data after $\sim$170$^{d}$ and 
suggesting positron escape. The Chandrasekhar mass models fit the
normally-luminous data 
better than does the sub-Chandrasekhar mass model, but the improvement is 
modest. 
As was discussed in MTL, the 
super-luminous model light curves are similar, both to other 
super-luminous SN models and to 
normally-luminous Chandrasekhar mass model light curves. We assert that the
16 
normally- and super-luminous SNe 
do not need to be differentiated into distinct sub-classes after 60$^{d}$.
We further assert that if the V band emission scales with the bolometric
emission 
to the level of a few tenths of a magnitude, then the late emission from all
16 SNe
 can be suitably explained with a 
single light curve featuring positron escape, and that Chandrasekhar mass, 
sub-Chandrasekhar mass, and merger explosions can all explain these data.

It is apparent from Figure 3 that the sub-luminous sub-class differs
considerably 
from the normally-/super-luminous sub-classes. MTL concluded that none of
the models 
they tested could explain the 60$^{d}$ -560$^{d}$ evolution of the 
light curves. One interpretation of that result is a rejection of both 
the Chandrasekhar mass and the sub-Chandrasekhar mass models tested. 
A different interpretation would be the failure of the assumption that 
the energy deposition rate scales with the V band 
during that epoch. Figure 5 
shows five sub-luminous models fit to the sub-luminous sub-class data. The
models 
are normalized to fit the data at a later epoch ($\sim$170$^{d}$). It is 
clear from the figure that none of the models can explain the V band 
data before 150$^{d}$.\footnote{The fifth panel (lowest) is an 
accretion-induced collapse model, which ejects 0.2 M$_{\odot}$ of material.
The 
model was created by scaling ONeMg to 0.2 M$_{\odot}$. This model 
is included upon the suggestion of Fryer et al. (1996,1999) that 
AIC's may explain sub-luminous SNe Ia. Even at 
this extremely low mass, the early escape of positrons is not enough to 
reproduce the steepness of the 65$^{d}$ -170$^{d}$ light curve.} 
After 150$^{d}$, the V band data can be 
suitably explained by all models if positron escape is assumed.\footnote{ 
It is important to note that the range of ionizations allowed lead to 
fairly thick curves for low-mass models, which exaggerates the visual 
impression of their ability 
to explain the data. A treatment of the level of ionization would reduce the
range to a single curve.} 
This latter interpretation implies that the energy deposition rate scaling 
with the V band occurs later for the sub-luminous sub-class than 
for the other sub-classes. For this interpretation to be correct, there 
 must be an explanation which is consistent with the spectral
observations, which suggest that sub-luminous SNe Ia enter the nebular 
phase earlier than normally-luminous or super-luminous SNe Ia (Mazzali et
al. 1997).

\subsection{BVRI Bands}

In the previous section, we fit model-generated energy deposition rates to 
V band data. That procedure relies upon the assertion that the V band 
traces the bolometric luminosity. For the normally-/super-luminous data-set,
the 
data are approximately fit with the models, suggesting a validation of 
that assertion. By contrast, for the sub-luminous data-set the models could 
not fit the data until a later epoch. 
All SNe Ia used in this study were also observed in 
the B, R and I bands. By independently fitting these data-sets to the 
model-generated energy deposition rates, we further explore the nature 
of the late-time emission from SNe Ia.

Shown in Figure 6 are the BVRI data-sets for SNe Ia fit to the model, W7. 
Within each sub-class the data has been normalized to 65$^{d}$. Each 
sub-class has then been 
independently fit to the model light curves. For all four
photometric bands, the normally-luminous and super-luminous light curves are
similar. 
The B band data for the normally-luminous and super-luminous SNe Ia deviate 
considerably from the model before 100$^{d}$, but by $\sim$170$^{d}$ 
they roughly scale with the energy deposition rate. 
The B band data for the sub-luminous sub-class 
differ from the normally-luminous and super-luminous sub-classes early, but
appear 
similar at late times. 
Perhaps the most striking results are in the R and I bands, 
where the three sub-classes are quite similar. We argue 
that the differences between the late {\it evolution} of the B and V band
data do 
not carry over to the longer wavelengths. 
The model fits initially deviate from the R and I band data for all three
sub-classes,  
but then converge upon the model curve by 
$\sim$170$^{d}$. The convergence in the I band is less certain due to 
scatter. 

It is interesting to note that the onset of the energy deposition 
scaling with the photometry ($\sim$170$^{d}$) occurs later than the 
transition from photospheric continuum emission to nebular emission 
($\sim$60$^{d}$) for 
all three SN Ia sub-classes and all four photometric bands. 
To determine whether this delay is explainable as a consequence of 
color evolution, we 
produce spectral templates for normally-/super-luminous and sub-luminous SNe
Ia. 
The templates have been created from a collection of SN Ia spectra;  
1994D (Patat et al. 1996; Turatto 2000; Filippenko 1997),  
1991T (Filippenko et al. 1992), 
1996X (Salvo et al. 2001), 
1991bg (Turatto et al. 1996; Ruiz -Lapuente et al. 1993), 
1987L (Ruiz -Lapuente et al. 1993), 1994ae (Bowers et al. 1997), 
1998de (Modjaz et al. 2000), 1998bu (Jha et al. 1999; Matheson 2000), 
1995D (Matheson 2000), 
1981B (Branch et al. 1983),
1984A (Branch 1987), 
1991F (Gomez \& Lopez 1995), 1992A (Kirshner et al. 1992).
We include all emission within the 3500\AA -9700\AA \hspace{1mm} wavelength
range, 
and ignore all emission outside of this range. Spectra that do not span that
range 
have been linearly interpolated from earlier and later spectra.\footnote{The
interpolation was performed in two steps. The shape of the interpolated
portion was 
determined from the two complete spectra adjacent in time. That shape was
then 
spliced to the existing portion.} The spectral sequences for a sampling of
each 
sub-class are shown in Figures 7 and 8.  
From these templates, we have determined the evolution of the ``filter 
efficiency'', or the fraction of the energy emission in 
the 3500\AA -9700\AA \hspace{1mm} wavelength range that is detected by each
band. 
The filter efficiencies for normally-luminous and super-luminous SNe Ia, 
shown in Figure 9, match the shape
of the residuals of the data-model fits shown in Figure 6 for the 
normally-/super-luminous data-set. The V band filter efficiency remains
nearly 
constant, in agreement with previous assertions that the V band scales with
the 
bolometric luminosity.\footnote{Contardo et al. (2000) arrived at a 
similar result.} Shown in Figure 10 are the
B,V,R and I band data fit to the model W7, after deriving bolometric 
corrections from these filter efficiencies. 
The data for all four bands are suitably explained by 
the energy deposition rates of the model if positron escape is 
assumed. Before $\sim$60$^{d}$, the B,V and I 
data vary dramatically from the 
energy deposition rates. This would be consistent with the decrease 
to negligible values of the photon diffusion time-scale, as suggested 
by Pinto \& Eastman (2000a). The correction lessened, but did not completely
remove the residuals from the B and V band data 
during the 60$^{d}$ -170$^{d}$ epoch. It is unclear whether the 
residuals remain due to a modest 
failure of this relatively crude bolometric correction calculation, 
or rather they remain because this treatment has ignored some 
physical phenomenon.

The application of these corrections essentially creates 
bolometric light curves from each photometric band. 
A single bolometric light curve for normally- and super-luminous 
SNe Ia was generated by combining the 
four curves shown in Figure 10 after weighting each curve according to 
the photometry errors. The resulting bolometric light curve can be 
compared with model-generated light curves. These comparisons, 
shown in Figure 11, exhibit the same tendencies that are seen in the V band;
the energy deposition rates for all 
four models are consistent with the data, and 
positron escape is suggested. These curves differ slightly from the
bolometric 
light curve derived by Contardo et al. (2000) from much of the same 
data. That study was primarily concerned with emission near the luminosity
peak 
and assumed constant bolometric corrections at late times ($\geq$
130$^{d}$). 
As we have shown that 
some evolution occurs during the 60$^{d}$ - 200$^{d}$ time period, we assert
that this bolometric light curve is more accurate at late times than the
light 
curve from Contardo et al. (2000). Although bolometric light curves 
include information from four photometric bands rather than simply 
from the V band, the 
V band is the best observed. Thus, we present the bolometric light curve 
primarily to demonstrate that conclusions derived from studies of the V band
shape are valid.

There are few nebular spectra for sub-luminous SNe Ia, so for this 
sub-class, the filter efficiencies are very crude and are 
undersampled after 150$^{d}$ (Figure 12). This is problematic, because 
as shown in 
Figure 6, the 100$^{d}$ -200$^{d}$ time-span is a critical transitional 
epoch. 
Nonetheless, as shown in Figure 13, the B,V,R and I band data are 
also fairly sparse and can 
be roughly explained with the model PDD54, after correcting for filter 
efficiency. All four bands fit the energy deposition rates better with 
the correction, as seen through comparison with Figure 6. 
There are large residuals in the V and I 
bands before $\sim$50$^{d}$. This is likely due to 
photon diffusion delays. 
It is evident in Figure 13 that the light curves of 
sub-luminous SNe Ia are consistent with model-generated energy deposition
rates 
which allow positron escape. More spectral and photometric 
data of sub-luminous SNe Ia need to be collected to further the 
understanding of this epoch. Due to the crudeness of the bolometric 
corrections, a bolometric light curve was not generated for the 
sub-luminous sub-class. 

The success of the bolometric corrections in reducing the 
residuals during the 60$^{d}$ -170$^{d}$ time-span
 suggests the existence of an epoch during which optical 
emission is instantaneously recycled, but the SN color 
continues to evolve. A possible explanation for this color evolution 
is the decay of $^{56}$Co to $^{56}$Fe. 
It is during the 65$^{d}$ -170$^{d}$ epoch that the daughter becomes the
dominant species in that decay. 
As shown by both Liu et al. (1997a,c) and Bowers et al. (1998), 
cobalt emission dominates the wavelength ranges observed with the R and I 
bands. As $^{56}$Co decays to $^{56}$Fe, these bands would fade faster 
than the energy deposition rate due to color evolution until late-times 
at which time most cobalt has decayed and iron and 
stable nickel dominate the spectrum in that wavelength range. In the V band 
wavelength range, both Co and Fe lines are present and the bolometric 
correction would change only slightly during the transition. 
The B band may brighten due to the increased emission from the many 
[FeII] lines from 446nm -456nm. This explanation can be tested 
through comparisons of model-generated spectra 
with observations made during this epoch. 

\section{Discussion}

Two tasks comprised this work. The first task was the 
compilation of a data-set of 
B,V,R,I photometry of type Ia SNe divided into sub-classes, and
development of bolometric correction factors for each band.
The second task was the generation of late-time light curves from SN Ia 
models and the fitting of these models to the compiled data-set. 
Certain findings can be derived from this study independent of 
the model-fitting, and are worthy of mention. We have 
demonstrated that there is order to the late light curves. The late 
B and V band light 
curves of sixteen normally- and super-luminous SNe 
follow a similar evolution, while those of six sub-luminous SNe 
follow a different evolution. By contrast, the late R and I band 
light curves of all 22 SNe Ia follow a similar evolution. 
The normally-/super-luminous sub-classes appear virtually identical although
they 
span more than half of the $\Delta$m$_{15}$(B) range of SNe in this 
study. The B and V band light curves of the sub-luminous SNe
appear to require a distinct sub-class.
Whether there are in fact SNe Ia that connect them to
normally-luminous objects is an important question. This question can 
only be answered by observational campaigns that focus upon monitoring 
SNe that span the range of $\Delta$m$_{15}$(B) values.

Comparisons of model-generated energy deposition rates with 
photometry in the B,V,R, and I bands reveal three epochs. There
is an  early epoch ($\leq$ 60$^{d}$), where energy
deposition/photometry  comparisons are invalid presumably due
to the time delay between  the deposition of energy and the
emergence of the resultant optical  emission. That epoch is
followed by an intermediate epoch  (60$^{d}$ -170$^{d}$) where
the bolometric light curve is fit by the  energy deposition
rate, but color evolution must be addressed for  individual
photometric bands to be fit by the energy deposition rate. 
This epoch is followed by a late epoch ($\geq$ 170$^{d}$) where
the  bolometric corrections are roughly constant and each
individual  photometric band can be fit by the energy
deposition rate if positron  escape is allowed for. These
epochs exist for both the normally-/super-luminous  sub-class,
and the sub-luminous sub-class. The bolometric  corrections
during the intermediate epoch are better defined for the 
normally-/super-luminous SNe Ia than for the sub-luminous SNe
Ia,   the latter corrections are preliminary. 

Collectively, the light curves of SNe Ia after 60$^{d}$ suggest that the 
interactions of the products of the $^{56}$Co $\rightarrow$ $^{56}$Fe 
decay with the ejecta can explain the energy deposition without any 
additional energy deposition source. Positrons are seen to escape the 
ejecta in quantity for all SNe Ia during the late epoch. These 
findings hold for Chandrasekhar mass, sub-Chandrasekhar mass, merger  
and AIC models equally. The model independence of these results does 
not imply that the late emission from SNe Ia cannot probe 
the progenitor. The challenge for NLTE radiation transport calculations 
is to reproduce the spectra during this epoch.

There are two ramifications of positron escape from SN Ia ejecta 
that warrant further discussion. First, nebular spectra have been used 
to estimate the $^{56}$Ni production in SNe Ia (Ruiz-Lapuente \& 
Filippenko 1996, Bowers et al. 1997) and to differentiate between 
Chandrasekhar mass and sub-Chandrasekhar mass 
models (Liu et al. 1997a,1997b,1998). In all of these 
studies, instantaneous positron energy loss 
was assumed. By 300$^{d}$, the 
energy deposition is dominated by positron slowing, and a substantial 
fraction of those positrons are expected to escape 
the ejecta in the radial scenario. 
Even positrons that do not escape the ejecta diffuse 
from the location of their creation. In light of this and other recent 
studies, nebular spectrum 
studies should be calculated with realistic
positron transport. Second, positrons that escape the SN Ia 
ejecta are thought to survive non-thermally on time-scales of 10$^{5}$ 
years, or more (Chan \& Lingenfelter 1993, 
Guessom, Ramaty \& Lingenfelter 1991). The collective positron 
contributions from SNe Ia, as inferred from extra-galactic SN rates, 
are sufficient to generate a large fraction of the 511 keV positron 
annihilation radiation 
observed by the CGRO/OSSE, SMM and TGRS gamma-ray telescopes (Milne et al. 
1999,2000). The degree of dominance is enough that accurate characterization
of positron annihilation radiation will trace recent SN Ia activity in 
the Galaxy. 

Light echoes have been detected from two SNe Ia, 1991T and 1998bu. 
In addition, 
Boffi et al. (1998) have identified SN 1989B as a light echo candidate.
Of all SNe Ia observed after 300$^{d}$, these three are the only SNe to 
remain significantly brighter than those in this work. As 
light echoes are currently being used as geometric distance indicators, 
this work suggests that the templates shown in this work may be used to 
locate light echoes (Sparks et al. 1999). 

There remain incomplete aspects to this study. More SNe Ia need to be 
monitored to reduce gaps in observations and determine the late 
behavior of more SNe Ia along the $\Delta$m$_{15}$(B) sequence. 
The 
exact nature of the transition from the onset of instantaneous 
recycling to the cessation of color evolution must be better observed 
spectrally and in each photometric band, particularly with 
sub-luminous SNe Ia. These unexplained aspects 
underscore the fact that much is to be learned about the physics of 
Type Ia SNe from late observations.

We thank P. H\H{o}flich for ongoing access to SN models and for discussion 
relating to radiation transport through SN ejecta. We thank 
M. Turratto and S. Jha for access to SN photometry and spectra. 
We also thank T. Matheson, R. Lopez, and P. Meikle for SN spectra.

\begin{deluxetable}{lccccc}
\footnotesize
\tablecaption{SN Ia model parameters. \label{table1} }
\tablewidth{0pt}
\tablehead{
\colhead{Model} & \colhead{Mode of } &
\colhead{M$_{\star}$} & \colhead{M$_{Ni}$ } &
\colhead{E$_{kin}$} & \colhead{Ref.}  \\
\colhead{Name} & \colhead{Explosion} & \colhead{[M$_{\odot}$]} &
\colhead{[M$_{\odot}$]} & 
}
\startdata
W7 & deflagration & 1.37 & 0.58 & 1.24 & 1 \nl
DD23C & delayed det. & 1.34 & 0.60 & 1.17 & 2 \nl
W7DT & late det. & 1.37 & 0.76 & 1.61 & 3 \nl
PDD54 & pul.del.det. & 1.40 & 0.17 & 1.02 & 4 \nl
HED6 & He-det. & 0.77 & 0.26 & 0.74 & 5 \nl
HED8 & He-det. & 0.96 & 0.51 & 1.00 & 5 \nl
HECD & He-det. & 1.07 & 0.72 & 1.35 & 6 \nl
ONeMg & AIC & 0.59 & 0.16 & 0.96 & 7 \nl
SmallFry & AIC & 0.20 & 0.05 & 0.32 & 8 \nl
DET2E2 & merger det.  & 1.40 & 0.62 & 1.33 & 5 \nl
DET2E6 & merger det.  & 1.80 & 0.62 & 1.33 & 5 \nl

\enddata

\tablenotetext{a}{REFERENCES.-(1)Nomoto et al. 1984; 
(2) H\H{o}flich et al. 1998; (3) Yamaoka et al. 1992; 
(4) H\H{o}flich, Khokhlov, \& Wheeler 1995, 
(5) H\H{o}flich \& Khokhlov 1996, (6) Kumagai 1997, 
(7) Nomoto et al. 1996; (8) Fryer, et al. 1999.}

\end{deluxetable}

\clearpage
\begin{figure}
\plotone{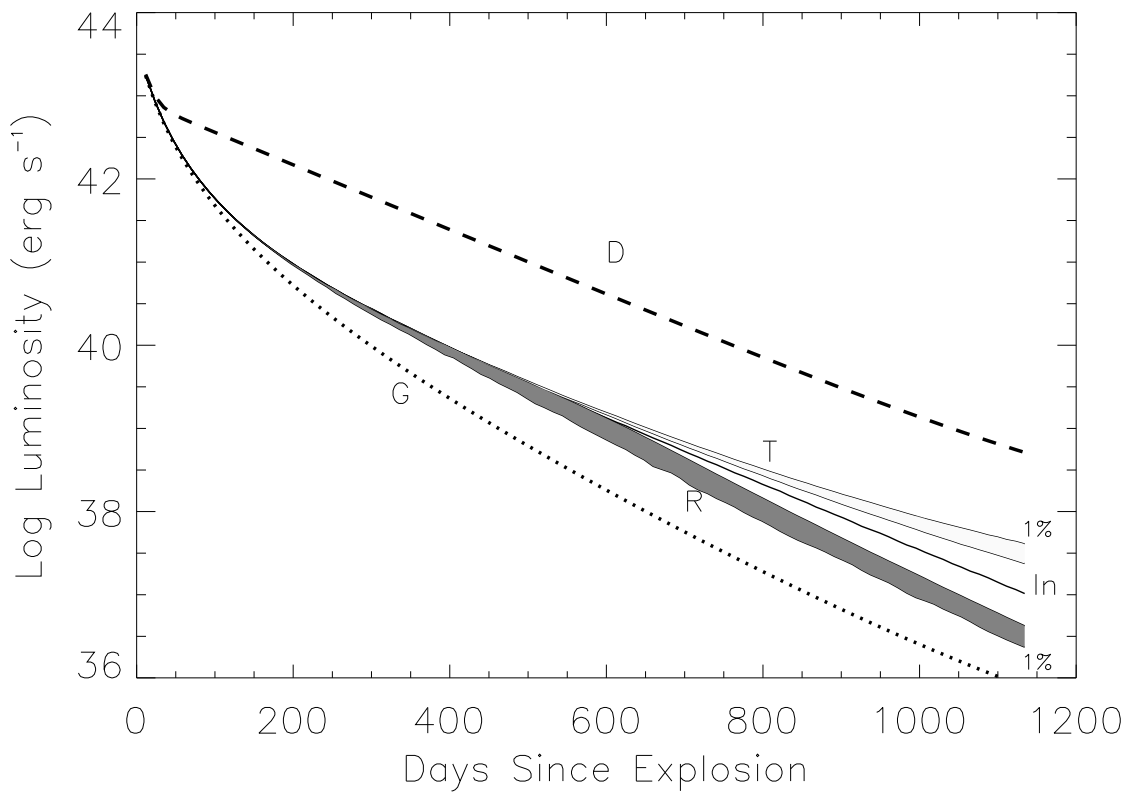}
\caption{A model generated bolometric light curve for the model HED8. The
dashed line (D) assumes instantaneous deposition of all decay energy. The
dotted line (G) uses the results of the gamma-ray energy deposition 
only and assumes
no deposition of positron energy. Between these two boundaries are the
results of the gamma-ray energy deposition coupled with instantaneous
positron deposition (thick line, In) and the range of curves for a  radial
field geometry (dark shading, R) and for a trapping geometry (light 
shading, T) as the electron ionization 
fraction varies from 0.01 $\leq \chi_{e} \leq$
3. \label{bol} }
\end{figure}

\clearpage
\begin{figure}
\plotone{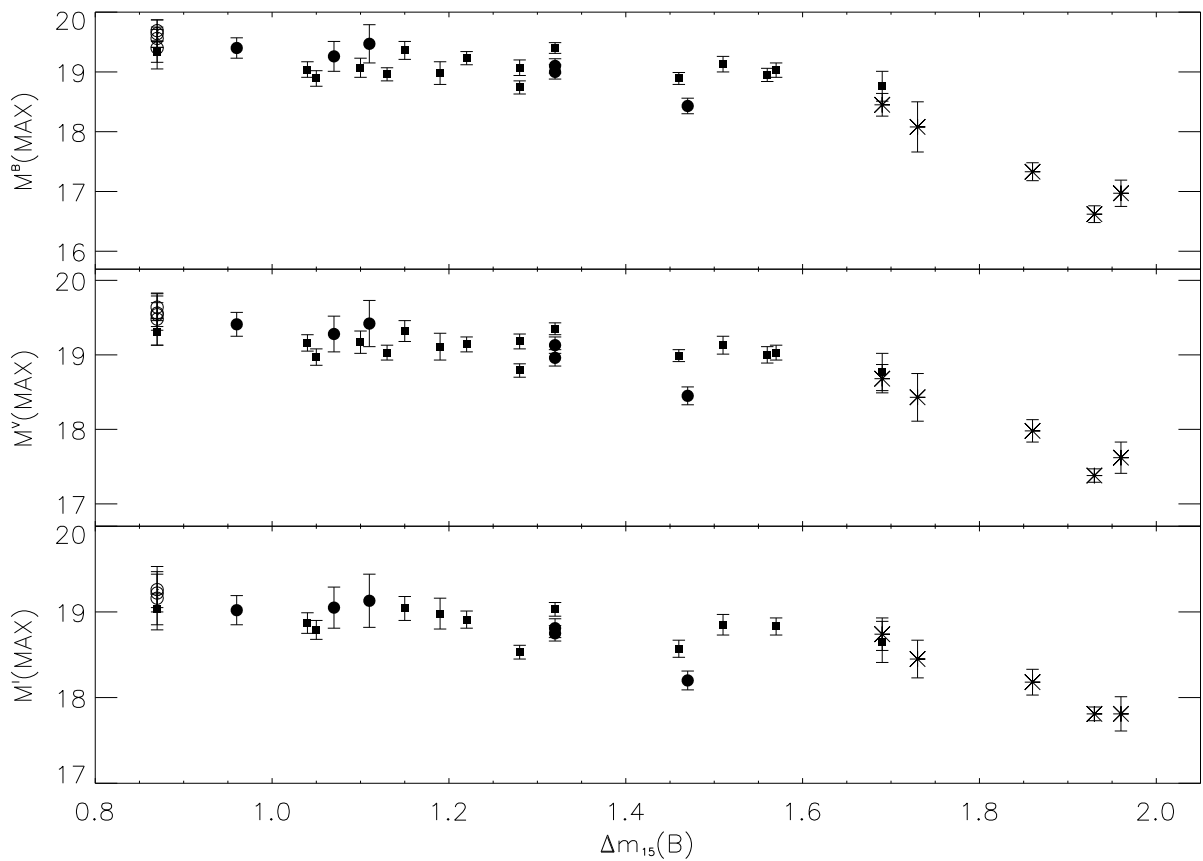}
\caption{The absolute B,V, \& I magnitudes of SNe Ia as a function 
of the $\Delta$m$_{15}$(B) parameter. All values taken from 
Hamuy et al. 1996, except SNe 1997cn (Turatto et al. 1998) and 
1998de (Modjaz et al. 2000). 
The normally-luminous, super-luminous and sub-luminous SNe Ia are shown 
as filled circles, open circles, and crosses, respectively, while 
other SNe not used in this study are shown as filled squares.
 All bands'
absolute magnitudes show a break around $\Delta$m$_{15}$(B)=1.6.
\label{mb15} }
\end{figure}

\clearpage

\begin{figure}
\plotone{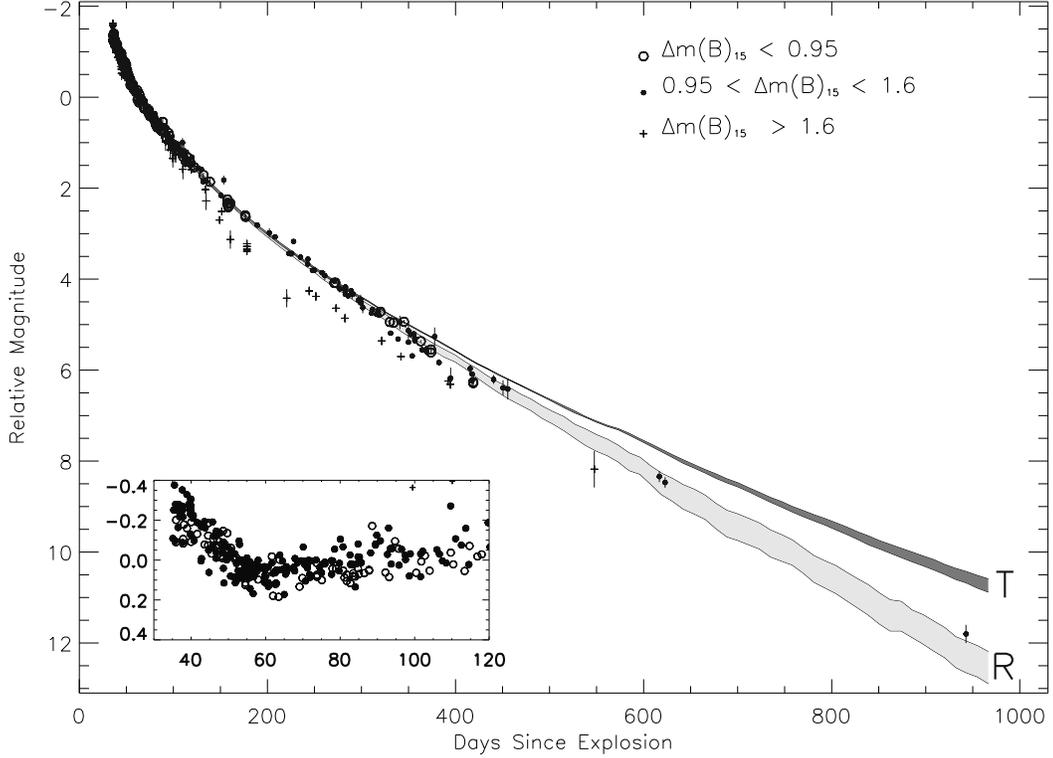}
\caption{The V band light curves of 22 SNe Ia fit with the 
energy deposition rates of the CM model, DD23C. Normally-luminous 
SNe Ia (filled circles) and super-luminous SNe Ia (open circles) 
appear to follow similar evolutions at late times, while 
sub-luminous SNe Ia (crosses) follow a different evolution. The data is 
better fit at late times by radial escape of positrons (R: light 
shading) than by positron trapping (T: dark shading). Inset: detailed 
30$^{d}$ -120$^{d}$ evolution of the {\bf N \& SP} sub-classes (see text).
SNe 1990O,1990T,1991ag,1992K,1992al,1992bc,1993H,1993ag from Hamuy 
et al. 1996. SNe 1994ae,1995D,1995E,1995al,1995bd from Riess et al. 
1997. Other SNe used, SN 1992A (Suntzeff 1996), 1990N (Lira 1998), 
1991T (Lira 1998; Schmidt et al. 1994), 1991bg (Leibundgut
et al. 1993; Turratto et al. 1996; Filippenko
et al. 1992), SN 1994D (Patat et al. 1996, Tanvir 1997, CAPP, Cappellaro 
1998), SN 1986G (Phillips et al. 1987, 1998, Cristiani 1992), 
SN 1997cn (Turatto et al. 1998), \& SN 1998de (Modjaz et al. 2000), 
SN 1996X (Salvo et al. 2001). SN 1995D also from Sadakane et al. 1996.
\label{reiss} }
\end{figure}

\begin{figure}
\plotone{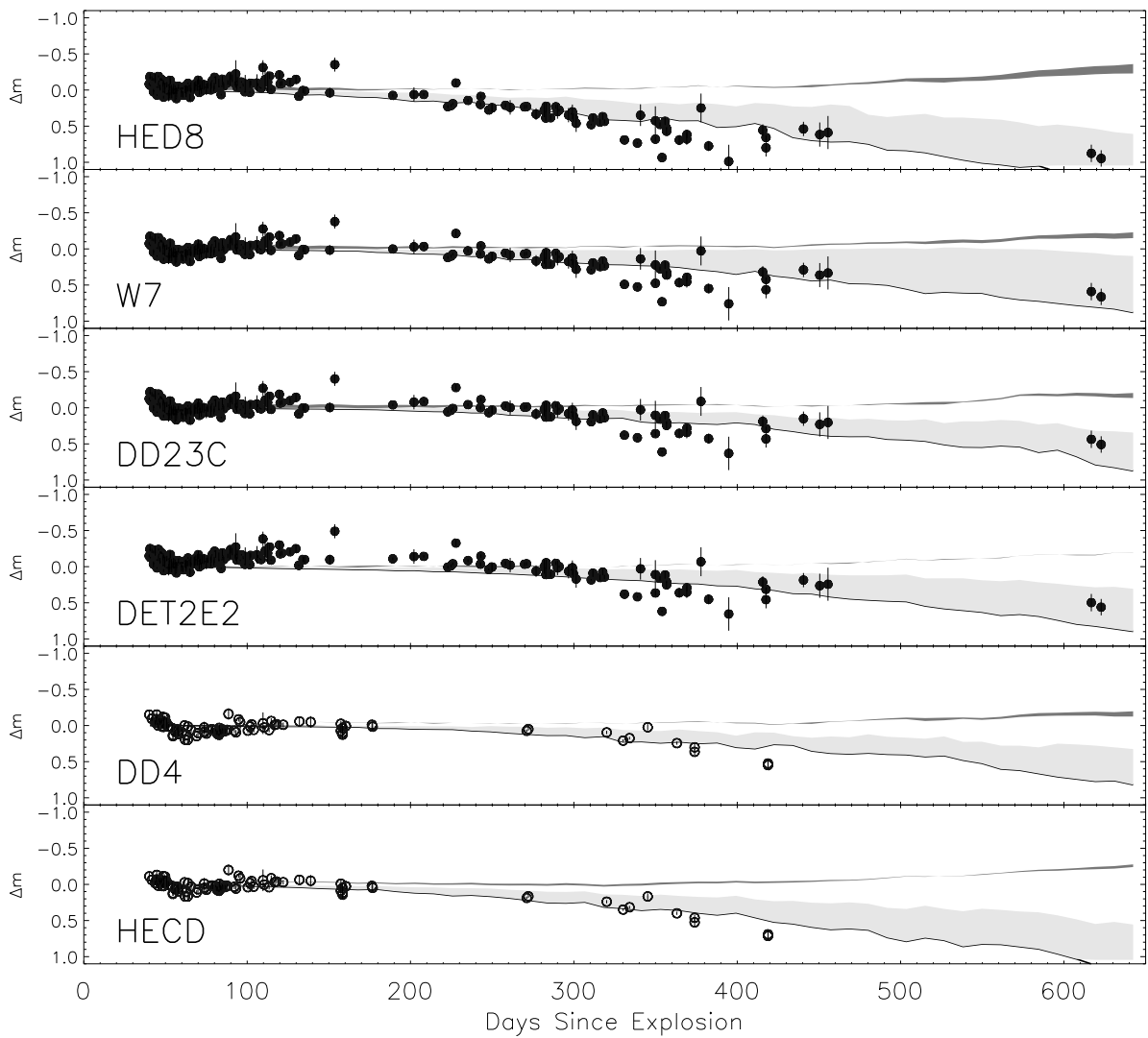}
\caption{The V band light curves of 16 normally-luminous
SNe Ia (upper four panels-filled circles) and 
super-luminous SNe Ia (lower two panels-open circles) 
fit with model-generated energy deposition rates from six SN Ia 
models, shown in the delta magnitude format. The four {\bf N} and 
two {\bf SP} models have been suggested to explain normally- and 
super-luminous SNe, respectively. 
The data is approximated by all six models after 170$^{d}$ 
if radial escape of positrons is assumed (lower band shaded curves).
The models' light-curves with positrons trapped (upper band shaded curves)
cannot give a good fit to the late-time data.
\label{vband1} }
\end{figure}

\begin{figure}
\plotone{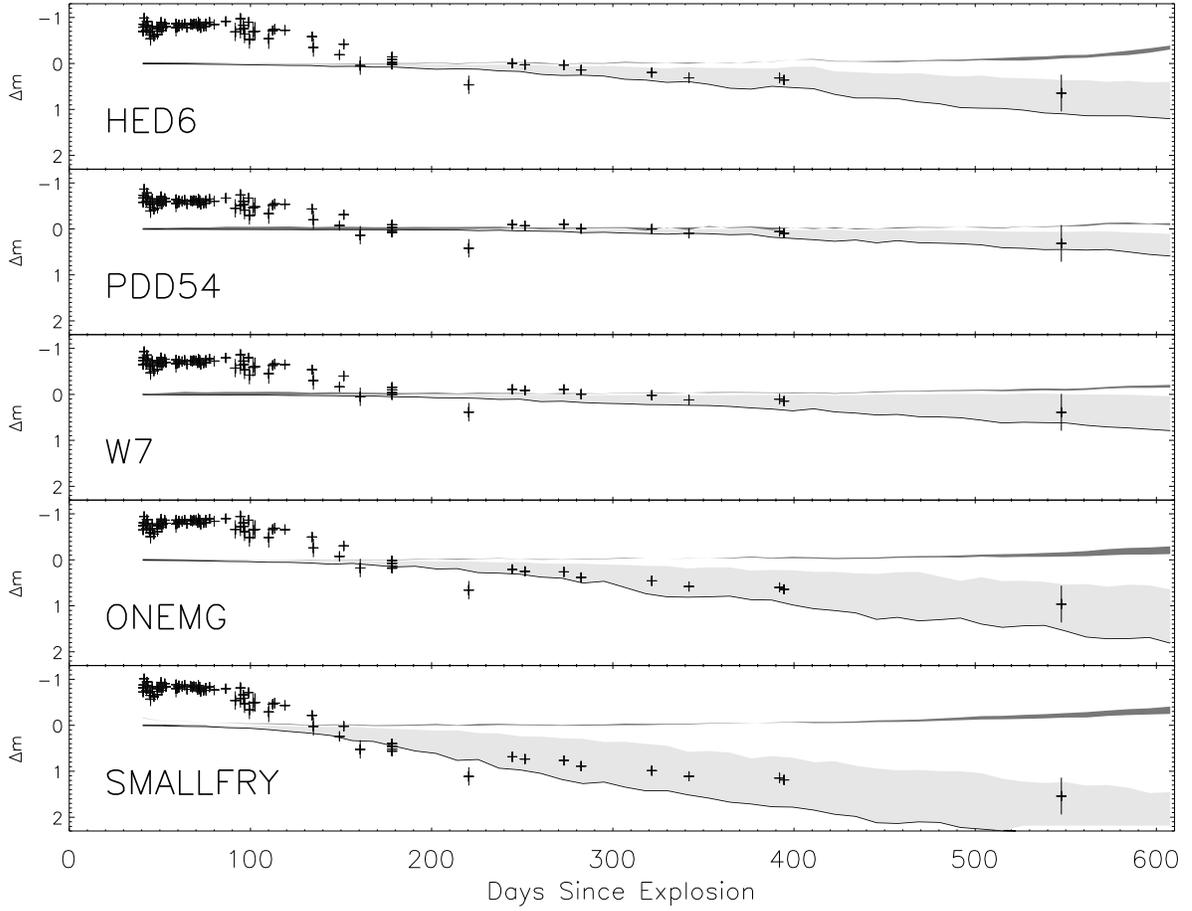}
\caption{The V band light curves of 6 sub-luminous SNe Ia 
fit with model-generated energy deposition rates from five SN Ia
models. Models and data have been transformed to the delta magnitude 
format. The data is consistent with all five models after 170$^{d}$
if radial escape of positrons is assumed.
\label{vband2} }
\end{figure}

\begin{figure}
\plotone{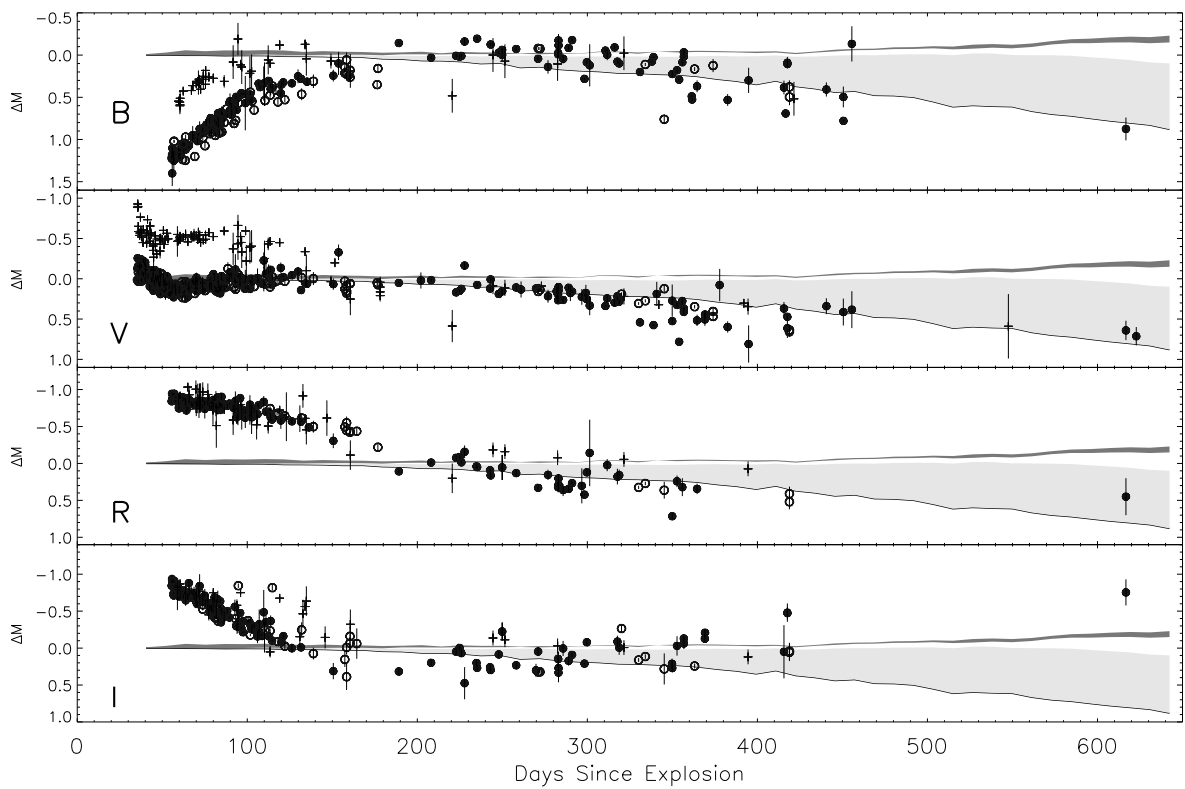}
\caption{The BVRI band light curves of 22 SNe Ia fit with 
model-generated energy deposition rates from the CM model, W7. 
Models and data have been transformed to the delta magnitude
format. The symbols of the data points are as in Fig. 4.
The {\bf sb} luminosity sub-class follows a different early 
evolution than the {\bf N \& SP} subclasses in the B \& V 
bands, but all three sub-classes are similar in the R \& I 
bands. For all SNe Ia, 
all four bands are approximated by the model-fit after 
170$^{d}$ if positron escape is assumed. 
 \label{nmhst} }
\end{figure}

\begin{figure}
\plotone{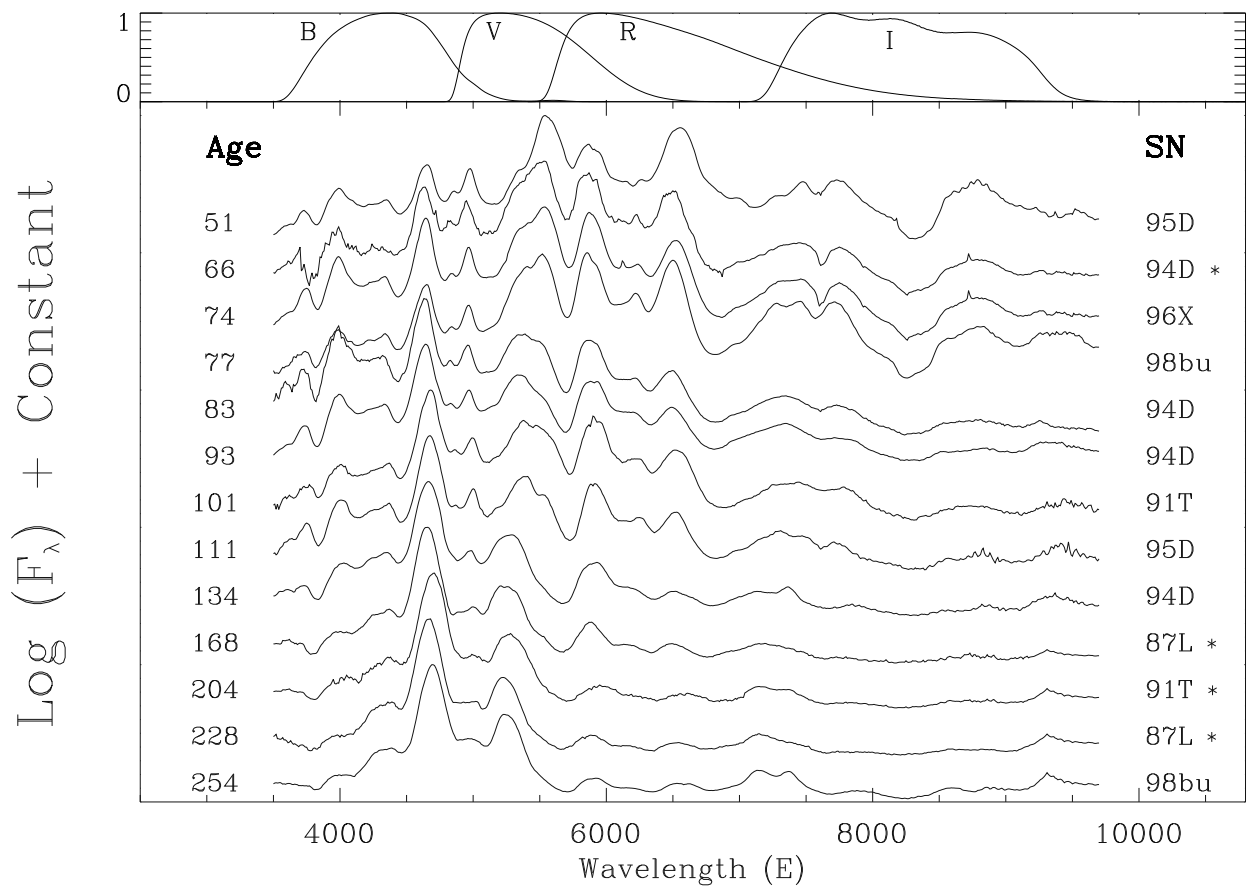}
\caption{A sequence of spectra of normally- and super-luminous SNe. 
All spectra have been offset by an arbitrary constant. Incomplete 
spectra are denoted with an asterisk (*). It is clear from these 
spectra that the emission within this wavelength range experiences 
a blueward shift. The transmission efficiencies for the B,V,R, and I 
filters are shown above the spectra for reference. \label{spseqnsp} }
\end{figure}

\begin{figure}
\plotone{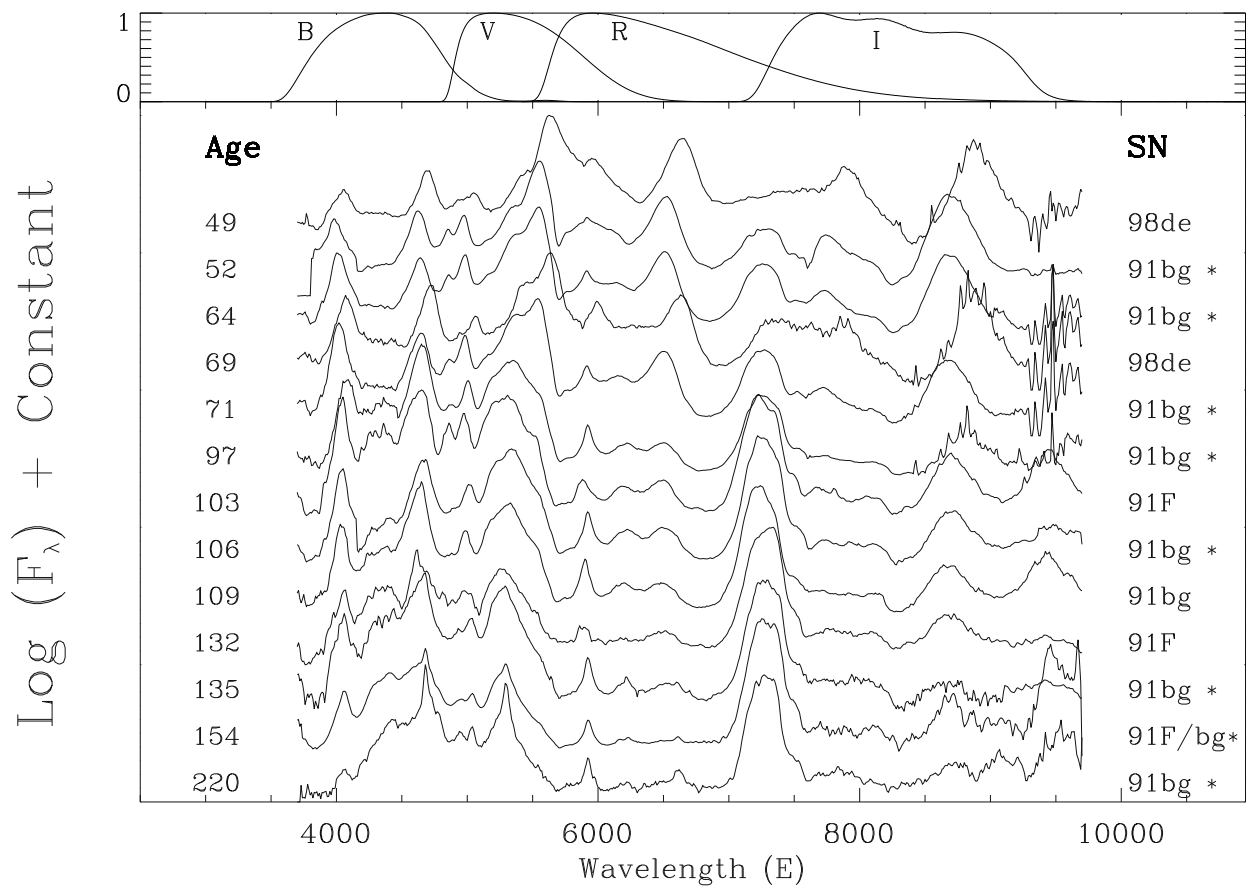}
\caption{A sequence of spectra of sub-luminous SNe.
As in Figure 7, 
all spectra have been offset by an arbitrary constant, and incomplete
spectra are denoted with an asterisk (*). This spectral sequence 
differs considerably from Figure 7, most notably due to the peak 
at 7300$\AA$, and the persistence of emission red-ward of 
8000$\AA$. The transmission efficiencies for the B,V,R, and I
filters are shown above the spectra for reference. \label{spseqsb} }
\end{figure}

\begin{figure}
\plotone{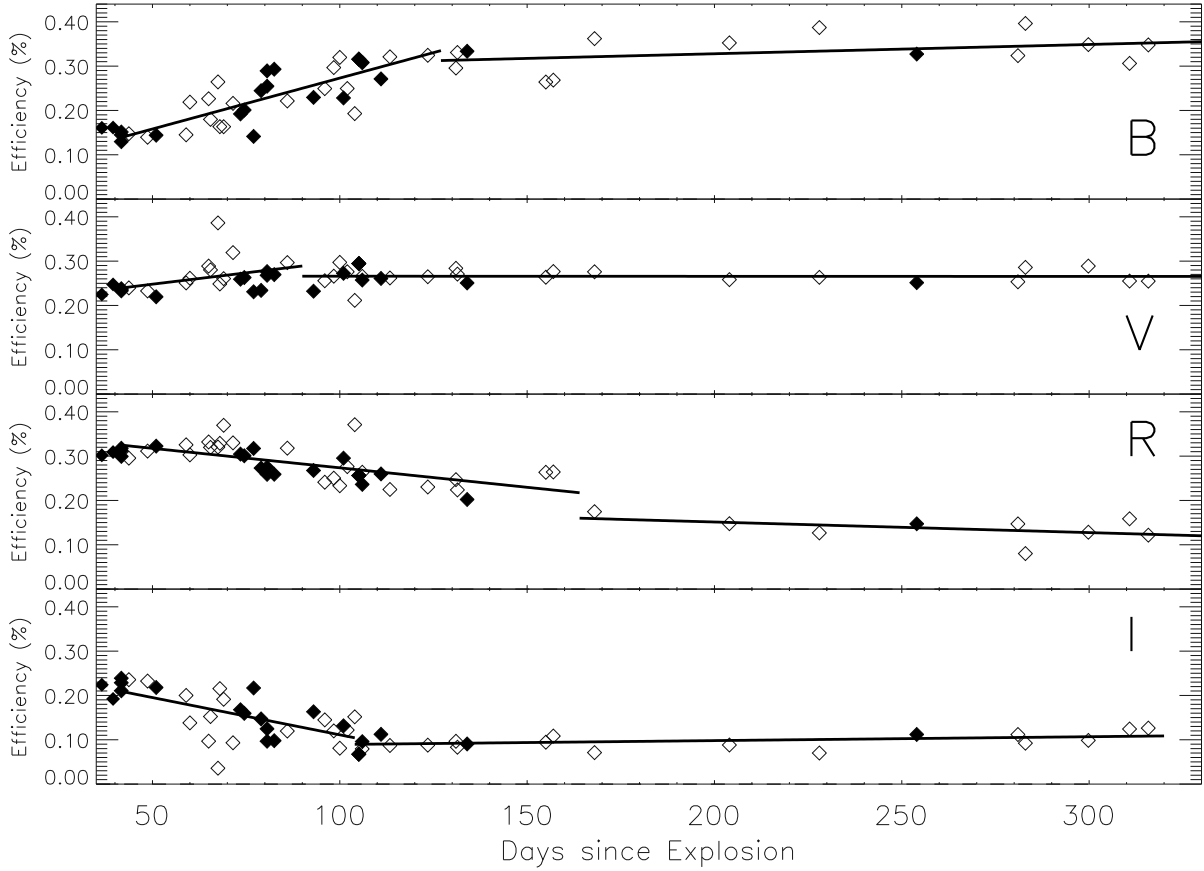}
\caption{The evolution of the fraction of the emission 
(3500\AA -9700\AA ) 
observable by the B,V,R and I bands for {\bf N} and 
{\bf SP} SNe Ia. Filled diamonds show spectra that spanned 
the wavelength range, open diamonds are incomplete spectra for 
which interpolation was required. \label{fcorrnsp} }
\end{figure}

\begin{figure}
\plotone{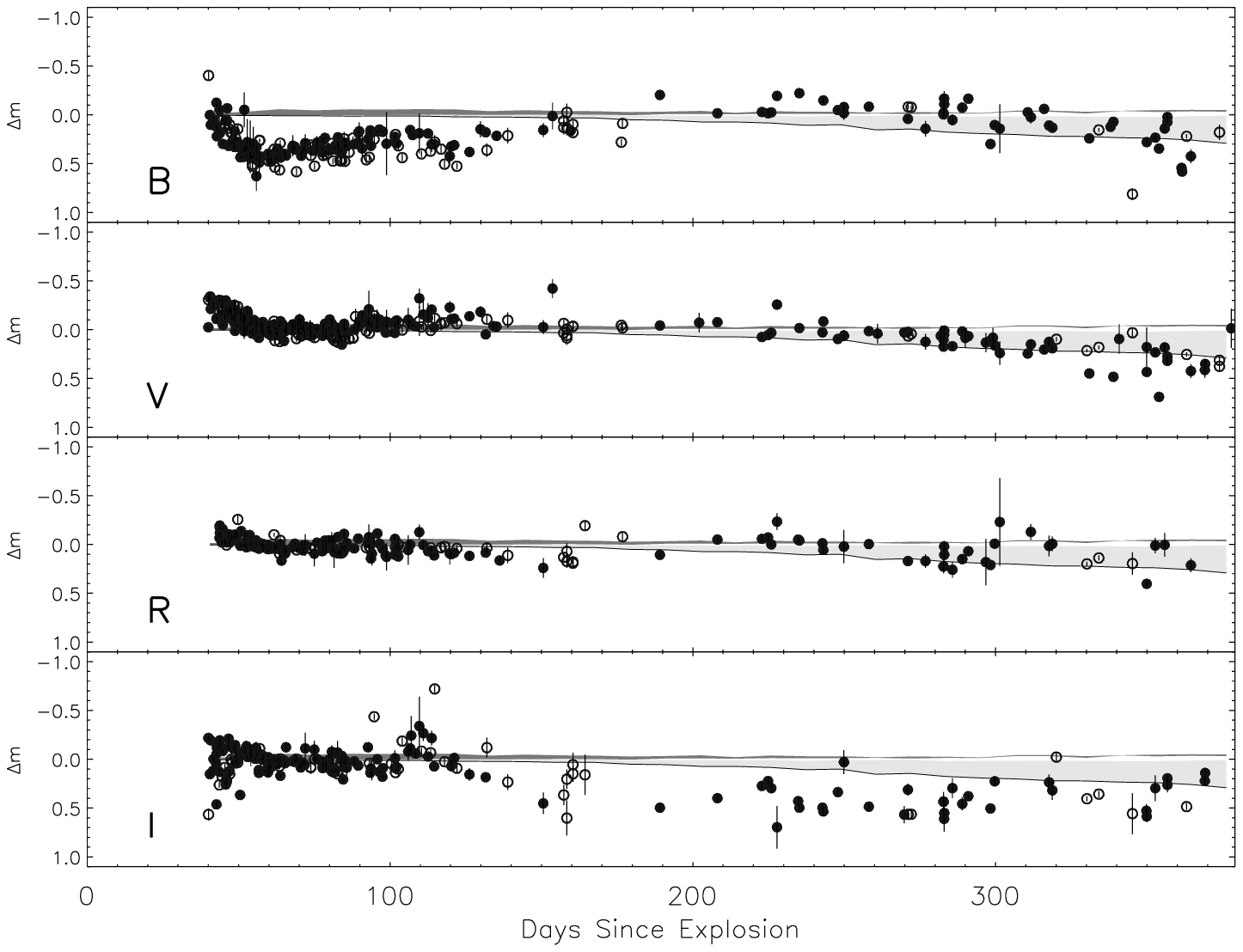}
\caption{Color-corrected BVRI band light curves of 16 {\bf N} 
and {\bf SP} SNe Ia fit with 
model-generated energy deposition rates from the CM model, W7.
As in Figure 6, the 
models and data have been transformed to the delta magnitude
format. \label{nmhst_nsp}}
\end{figure}

\begin{figure}
\plotone{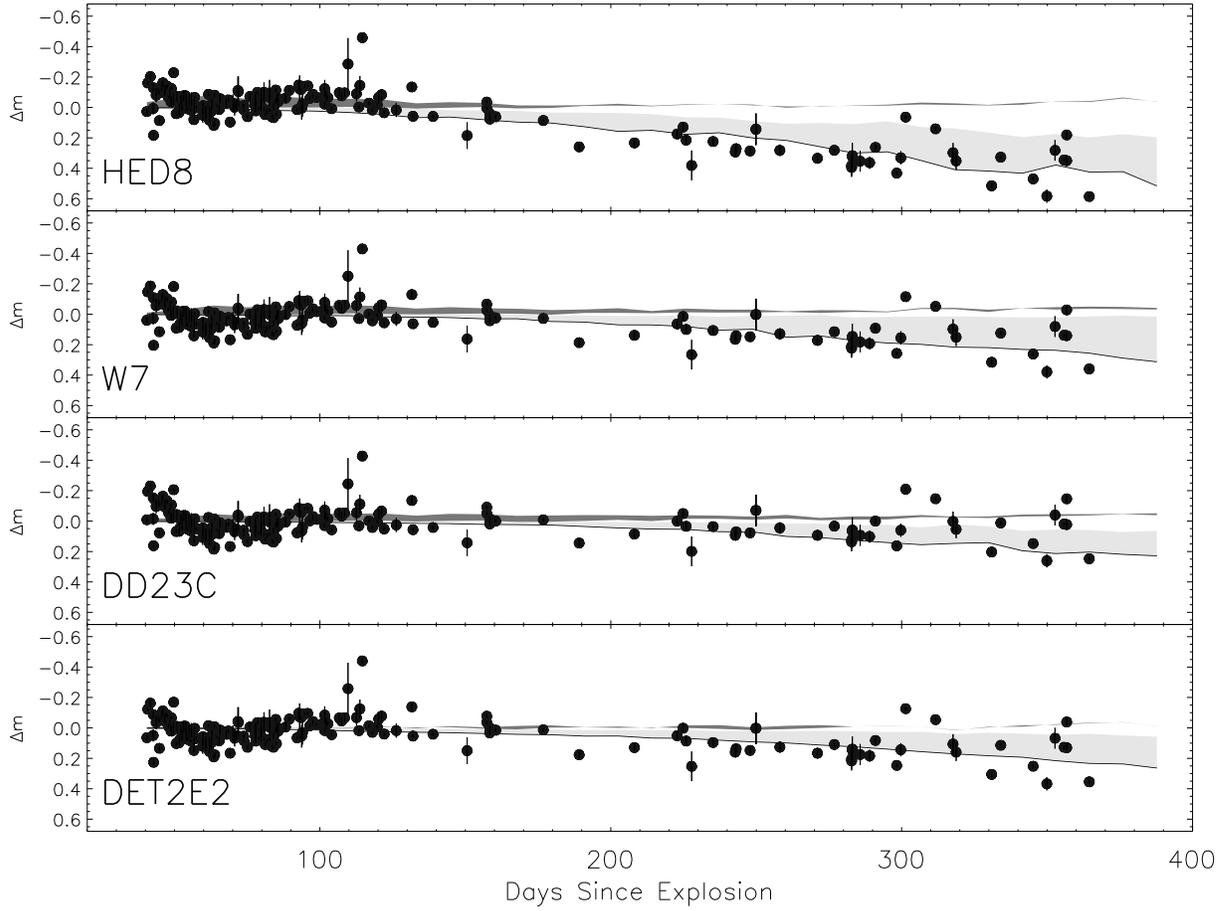}
\caption{Model-generated energy deposition rates for four 
SN Ia models compared with calculated bolometric light 
curves for 16 normally- and super-luminous SNe Ia.
\label{nmhst_mybol} }
\end{figure}

\begin{figure}
\plotone{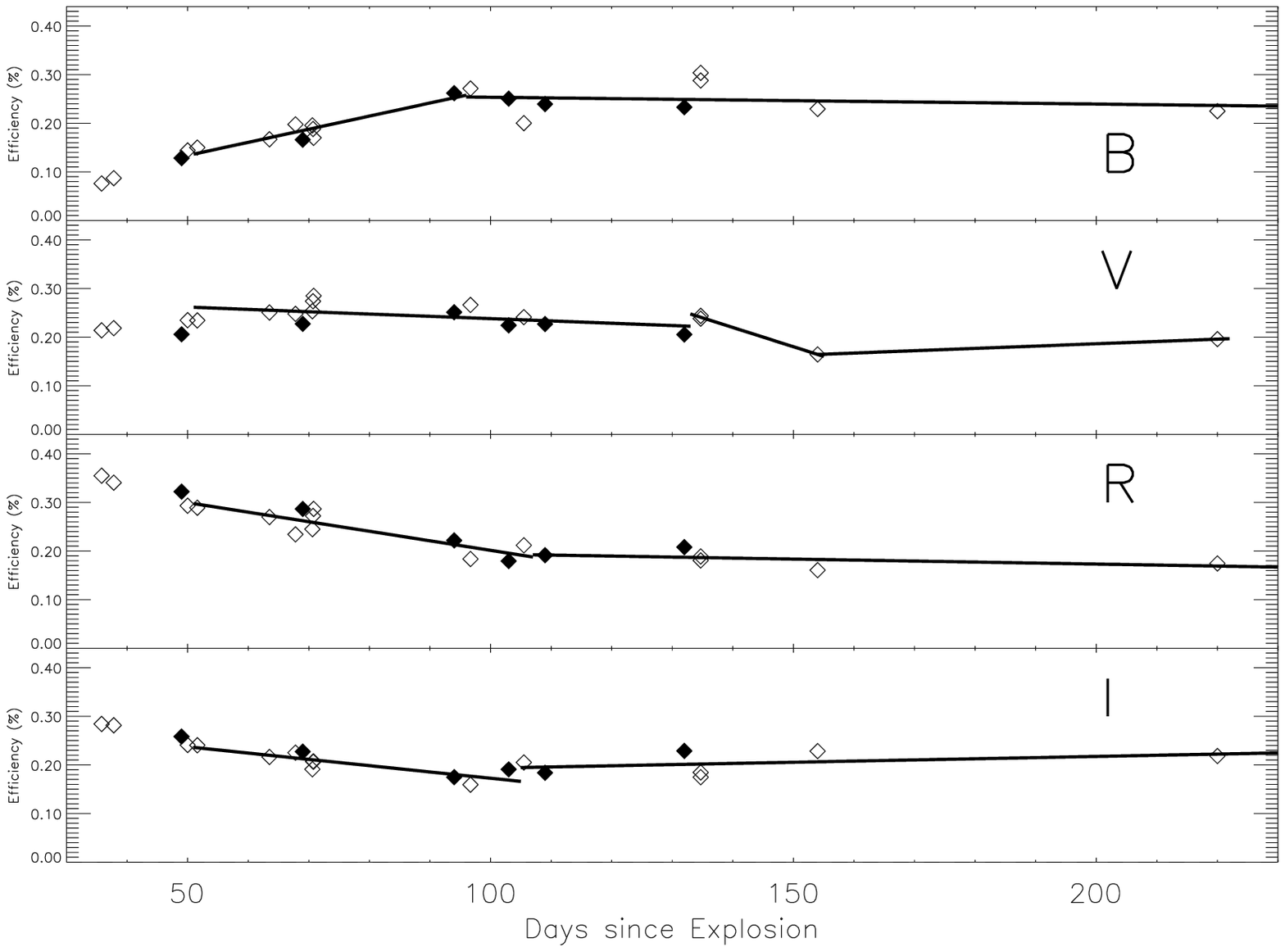}
\caption{The evolution of the fraction of the emission 
(3500\AA -9700\AA ) 
observable by the B,V,R and I bands for 
{\bf sb} SNe Ia. As in Figure 8, Filled diamonds show spectra that spanned
the wavelength range, open diamonds are incomplete spectra for
which interpolation was required. \label{fcorrsb} }
\end{figure}

\begin{figure}
\plotone{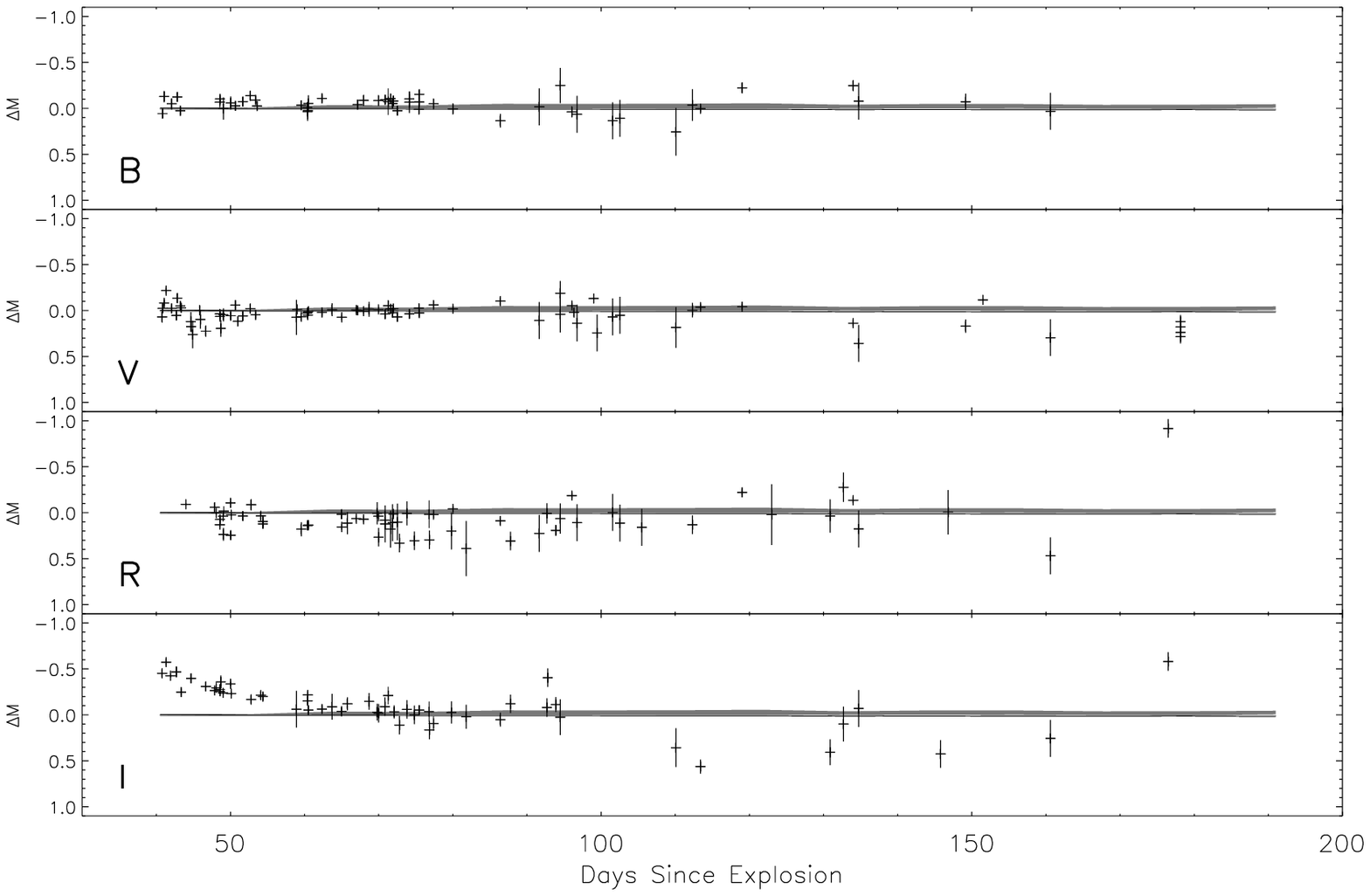}
\caption{Color-corrected BVRI band light curves of 6 {\bf sb} 
SNe Ia fit with
model-generated energy deposition rates from the CM model, PDD54.
As in Figures 6 \& 9, the
models and data have been transformed to the delta magnitude
format. \label{nmhst_sb}}
\end{figure}

\end{document}